# Ultra-narrowband terahertz circular dichroism driven by planar metasurface supporting chiral quasi bound states in continuum

Jitao Li [a,1], Zhen Yue [a,1], Jie Li [a,1], Chenglong Zheng [a,1], Dingyu Yang [b], Silei Wang [a], Mengyao Li [a], Yating Zhang [a, *], Jianquan Yao [a, **]

[a]Key Laboratory of Opto-Electronics Information Technology (Tianjin University), Ministry of Education, School of Precision Instruments and Opto-Electronics Engineering, Tianjin University, Tianjin, 300072, China

[b]Chengdu University of Information Technology, Chengdu, 610225, China

Corresponding author. E-mail:*yating@tju.edu.cn (Y. Zhang), **jqyao@tju.edu.cn (J. Yao).

[1]These authors contribute equally to this work.

**ABSTRACT**

Terahertz (THz) chirality pursues customizable manipulation from narrowband to broadband. While conventional THz chirality is restricted by non-negligible linewidth and unable to handle narrowband well. Recently, the concept "quasi bound states in continuum" (quasi-BIC) is introduced to optics resonance system whose the quality factor can be extremely high with the ultra-low radiative loss, thus providing a conceptual feasibility for wave control with ultra-narrow linewidth. Herein, we construct quasi-BIC in a planar all-silicon THz metasurface with in-plane $C_2$ and mirror symmetries breaking. Such system not only exposes the symmetry-protected BIC, but also exposes the parameter-tuned BIC assigned to single resonance type. An extremely narrow linewidth (below 0.06 GHz) with high quality factor ($10^4$ level) is obtained at quasi-BIC frequency, which achieves the ultra-narrowband THz chirality.

## 1. INTRODUCTION

The spin-selective asymmetric transmission of light waves can be applied to many fields such as imaging, communication and information [1-2]. Such function has been achieved in the conventional chiral metasurfaces [3-8]. The ideal terahertz (THz) chirality pursues customizable manipulation from narrowband to broadband. Nevertheless, the resonance response of the conventional chiral THz metasurface with a low quality (Q) factor (weak interaction) suggests a non-negligible bandwidth in spectrum. Obviously, the traditional principle is not applicable to narrowband THz chirality, thus it does make sense to explore an advanced theory to enhance interaction between THz waves and metasurface.

Bound states in continuum (BIC), a strong interaction with an infinite Q factor and zero leakage, has been proposed in metasurfaces recently [9-12]. The electromagnetic wave with frequency coexists with extended state in the radiating continuum but remains perfectly localized without radiation [13-14]. Therefore, BIC cannot be accessed directly, while a quasi-BIC with radiation leakage can be obtained by some physical changes to the structural parameters of metasurface [15]. Quasi-BIC can be tuned to a high Q factor with an ultra-narrow radiation linewidth [16-17], thus providing a conceptual feasibility for generating an ultra-narrowband electromagnetic wave. Recently, combined with a pair of dielectric pillars with different heights and a certain angle, the out-of-plane BIC breaking has demonstrated a great effect on chiral manipulation of narrow linewidth electromagnetic waves [18]. By constructing a pair of tightly stacked interfaces, the good chiral manipulation of narrow linewidth electromagnetic wave based on in-plane BIC breaking is also shown [19]. Nevertheless, previous reports relied on the non-planar structure that the mirror symmetry with respect to a plane perpendicular to the z-axis was destroyed, causing the inconsistent size along z-direction. Therefore, it becomes in practice difficulty to manufacture these structure by conventional etching method with two-dimensional pattern. Obviously, the size consistency at z-direction is more conducive to the actual preparation, which requires the unit cell to retain planar. However, reports on chiral quasi-BIC supported by planar metasurface remain lacking, thus requiring to be investigated.

In this work, a planar all-silicon THz metasurface supporting chiral quasi-BIC is proposed here. The meta-atom is a rectangular block with a square opening offset from the center. In-plane $C_2$ and mirror symmetries of such structure have been broken, but the mirror symmetry with respect to a plane perpendicular to the z-axis is preserved. Interestingly, different from previous reports where the system only expose symmetry-protected BICs [18-19], such system we reported also exposes the parameter-tuned BIC assigned to single resonance type. The results show that the circular dichroism and radiation linewidth of the THz wave are modulated by quasi-BIC mode. The metasurface successfully achieves ultra-narrowband THz chirality with linewidth below 0.06 GHz and Q factor of $10^4$ level. Overall, this work provides a new perspective for chiral quasi-BIC.

## 2. METHOD
### 2.1 Model design

The periodic planer photonic structures have a series of Bloch resonance modes with different frequencies and wave vectors forming photonic bands. The radiative polarization states of these modes on an arbitrary band could be mapped to the Brillouin zone (BZ), that is, the polarization field in momentum space is defined. The polarization state of the nonradiative at-$\Gamma$ BIC is uncertainty, i.e., polarization singularity, but a specific structural perturbation can turn it into a radiative quasi-BIC with a uniquely determined polarization state, and the reason was recently identified as the creation of nonzero multipole moments in the structure plane [20-23]. The physical connection between radiation polarization state in BZ and structural perturbation has been explored in detail by Jian Zi *et al*. [23], which can guide us to design chiral quasi-BICs located at the $\Gamma$ point (corresponding to normal incident light). Next, we demonstrate the design process of the at-$\Gamma$ chiral quasi-BIC, starting with the at-$\Gamma$ BIC. The finite element method solving Maxwell's equations was employed to complete all numerical simulation.

Take a symmetry-protected BIC supported by a rectangular structure as an example, where the eigen-field with bound mode are shown in Fig. 1a to simply confirm BIC, and more details for BIC confirmation are placed in Section 3. Fig. 1b-d show the distribution of radiation polarization states in the BZ for unit cells with different symmetries. The polarization state at each wave vector point was plotted by monitoring the electric field vector of the radiation leakage field of the resonant mode. The distribution symmetry of radiation polarization states in the BZ is closely related to the geometrical symmetry of the unit cell in real space. For example, for an unit cell with $C_2$ symmetry, the radiation polarization state distribution in BZ also has $C_2$ symmetry, as shown in Fig. 1b. The radiation polarization state at the off-Γ point in the BZ is almost linear without ellipticity, suggesting that it is impossible to excite the circular dichroism of such structure. Next, let the Γ point enter into the quasi-BIC with a radiation polarization state, and the most intuitive method is to introduce an opening to the central axis of the structure, as shown in Fig. 1c. At this time, the $C_2$ symmetry of the structure is broken but the in-plane mirror symmetry with respect to y-axis is preserved, so the polarization state distribution in BZ is also mirror symmetry with respect to $k_y$ ($k_y$ // y-axis). Nevertheless, the radiation polarization state at Γ point is linear, which means that the circular polarization modes with different helicity at Γ point are still fully coupled, so that circular dichroism is also not realized. Even so, the limitation on ellipticity at off-Γ point has been broken. It is easy to be found from Fig. 1c that changing the wave vector along the route of $k_x \neq 0$ can obtain elliptically polarized modes (e.g. point N), which give possibility to achieve circular dichroism --- move the Γ point to N with nonzero ellipticity. As mentioned above, the distribution symmetry of polarization states in BZ is closely related to the symmetry of unit cell in real space. Obviously, once the Γ point is moved to N, the polarization state distribution in BZ will loss both $C_2$ symmetry and in-plane mirror symmetry, which suggests, in real space, to design the structure as neither $C_2$ symmetry nor in-plane mirror symmetry. As shown in Fig. 1d, the opening position is moved to further break the in-plane mirror symmetry compared to that in Fig. 1c. The radiation polarization state of this structure at the Γ point is no longer linear, but produces circular polarization modes decoupling, thereby an at-Γ chiral quasi BIC is constructed.

The structural parameters can be summarized as Fig. 1e. The planar unit cell is based on a rectangular silicon block with a square opening (silicon dielectric constant $\varepsilon$=11.9 and conductivity less than 0.03 S/m [24]). The period constant of unit cell is Px=Py=160 μm, with the length of the rectangular block $l$=90 μm, the width $w$=60 μm and the height $h$=50 μm. The side length of the square opening is $g$=8-20 μm, and its position is away from the center of the rectangular block as $d$=18 μm. An appropriate opening position $d$ is selected to destroy the in-plane (in x-y plane) mirror symmetry of the structure ($C_2$ symmetry is also destroyed synchronously), but the mirror symmetry with respect to a plane perpendicular to the z-axis is still retained. The spin-selective asymmetric propagation with the quasi-BIC feature at Γ point will be obtained. In detail, a port preferentially transmits cross-polarization wave of incoming wave with a certain helicity, and reflects co-polarization wave of incoming wave with opposite helicity, as schematically shown in Fig. 1f. In this work, we define the left-handed circular polarization (LCP) and right-handed circular polarization (RCP) waves as the waves that the electric field vectors rotate with counterclockwise and clockwise along the light propagation direction, respectively. Meanwhile, not only for chiral quasi-BIC with symmetry-protected type, we will soon see that other types of chiral quasi-BIC may also be exposed in this structure (see Section 3).

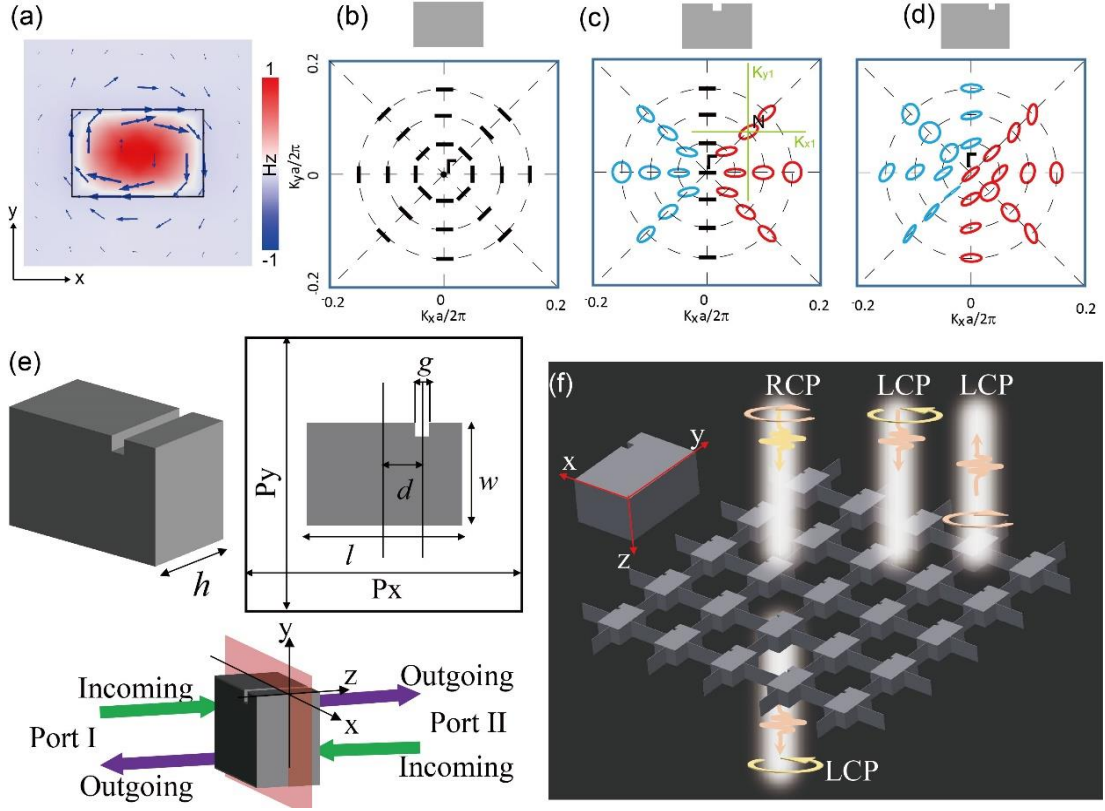

Fig. 1. (a) The schematic egien magnetic field distribution of an at-Γ BIC (1.09 THz, see Fig. 3b) supported by the planar rectangular structure, where blue arrows mean electric field vector. (b-c) The schematic view of the distribution of radiation polarization states in the BZ for different unit cells: (b) with $C_2$ symmetry, (c) without $C_2$ symmetry but keeping in-plane mirror symmetry, (d) both $C_2$ and in-plane mirror symmetries breaking. The black short lines in (b-c) correspond to the liner polarization states. The elliptical rings with blue and red in (c-d) mean the elliptical polarization states as counterclockwise and clockwise helicity. (e) The structural parameters for chiral quasi-BIC unit cell. This structure has no in-plane (in x-y plane) $C_2$ symmetry and in-plane mirror symmetry. A cross-sectional plane perpendicular to the z-axis exists in this planar structure making the elements on two sides of z-axis mirror symmetry. (f) The schematic function for chiral quasi-BIC metasurface. It should be noted that in schematic (f) unit cells are connected by slender strips as slender as possible, which does not affect the optical properties of the unit cell, and this design aims to provide a suggestion for possible realistic preparation.

## 2.2 Scattering matrix

Temporal coupled-mode theory (CMT) has been widely applied to provide a simple analytical description for optical resonance objects weakly coupled to incoming and outgoing ports. Temporal CMT works well in the weak-coupling regime; in practice, it is nearly exact when Q>30, which is the case considered in this work. Importantly, temporal CMT is a phenomenological theory applied to describe resonance results and does not consider the source of resonance. No matter what type of BIC is, radiation will occur with the formation of quasi-BIC. At this time, an essential element is the coupling between the resonator and ports. Therefore, temporal CMT is universally applicable for discussing the scattering properties of such structure [25]. By employing temporal CMT, the analytical expressions of scattering matrix can be deduced as Eq. 1. The reflection and transmission amplitudes possess the form of Fano linetype function with a center frequency $\omega_0$ and a radiative rate $\gamma_0$, and written as: $|(\Lambda+i\Gamma)/[i(\omega-\omega_0)+\gamma_0]|$, where $\Lambda$, $\Gamma$ are real numbers. In detail, considering incoming wave on one port, e.g. port I, the reflection and transmission coefficients read as (see **S1** in **Supplementary Information** for details):

$$\begin{bmatrix} r_{RR} & r_{RL} \\ r_{LR} & r_{LL} \end{bmatrix} = \begin{bmatrix} \dfrac{\Lambda_1+i\Gamma_1}{i(\omega-\omega_0)+\gamma_0} & \dfrac{\Lambda_2+i\Gamma_2}{i(\omega-\omega_0)+\gamma_0} \\ \dfrac{\Lambda_2+i\Gamma_2}{i(\omega-\omega_0)+\gamma_0} & \dfrac{\Lambda_3+i\Gamma_3}{i(\omega-\omega_0)+\gamma_0} \end{bmatrix}, \quad (1a)$$

$$\begin{bmatrix} t_{RR} & t_{RL} \\ t_{LR} & t_{LL} \end{bmatrix} = \begin{bmatrix} \dfrac{\Lambda_4+i\Gamma_4}{i(\omega-\omega_0)+\gamma_0} & \dfrac{\Lambda_5+i\Gamma_5}{i(\omega-\omega_0)+\gamma_0} \\ \dfrac{\Lambda_6+i\Gamma_6}{i(\omega-\omega_0)+\gamma_0} & \dfrac{\Lambda_4+i\Gamma_4}{i(\omega-\omega_0)+\gamma_0} \end{bmatrix}, \quad (1b)$$

where $\omega_0$ and $\gamma_0$ mean the quasi-BIC frequency and radiative loss, respectively. The $r_{jk}$ and $t_{jk}$ ($j,k \in \{L,R\}$, L means LCP, and R means RCP), respectively mean reflection and transmission coefficients of j-component under k-component wave incident on port I. The Eq. 1 qualitatively describes the scattering properties of metasurface supporting quasi-BIC. It can be seen from Eq. 1 that the reflection properties are $r_{RR} \neq r_{LL}$ and $r_{RL}=r_{LR}$, and transmission properties are $t_{RR}=t_{LL}$ and $t_{RL} \neq t_{LR}$. Moreover, the term $\Lambda+i\Gamma$ in Eq. 1 is expressed as:

$$\Lambda_j + i\Gamma_j = [\gamma_0 A_j + (\omega-\omega_0)B_j] + i[\gamma_0 C_j + (\omega-\omega_0)D_j], j \in \{1,2,3,4,5,6\}, \quad (2)$$

where real numbers $A_j$, $B_j$, $C_j$ and $D_j$ are related to the achiral background scattering and coupling between resonator and ports. Substituting Eq. 2 into Eq. 1, both reflection and transmission amplitudes can be expressed as a general form:

$$|r,t|_{Fano} = \left| \dfrac{[\gamma_0 A_j + (\omega-\omega_0)B_j] + i[\gamma_0 C_j + (\omega-\omega_0)D_j]}{i(\omega-\omega_0)+\gamma_0} \right|, j \in \{1,2,3,4,5,6\}. \quad (3)$$

Once the structure is determined, real numbers $A_j$, $B_j$, $C_j$ and $D_j$ will be determined.

## 3. RESULTS AND DISCUSSION

Considering the circularly-polarized THz waves incident on port I of chiral structure ($d$=18 μm, $g$=12 μm), the transmission spectra with Fano linetype in Fig. 2a show $t_{RR}=t_{LL}$ and $t_{RL} \neq t_{LR}$, which is consistent with the trend described in Eq. 1b. It can be seen from transmission spectra that this structure exposes three quasi-BIC frequencies in the range of 1.1-1.5 THz. For all quasi-BIC frequencies, spin-decoupling occurs and chiral transmission takes effect (Fig. 2b). Transmission circular dichroism (CD) defined as $|t_R|^2-|t_L|^2$ (subscript "L/R" means LCP/RCP incoming wave) will be used to appraise spin-selective asymmetric propagation. The CD shows that the frequencies on which chiral transmissions depend have a very narrow linewidth (defined as full width at half maximum). It should be noted that the CD of the three chiral quasi-BICS is not fixed, which can be changed by adjusting parameters such as opening size and position.

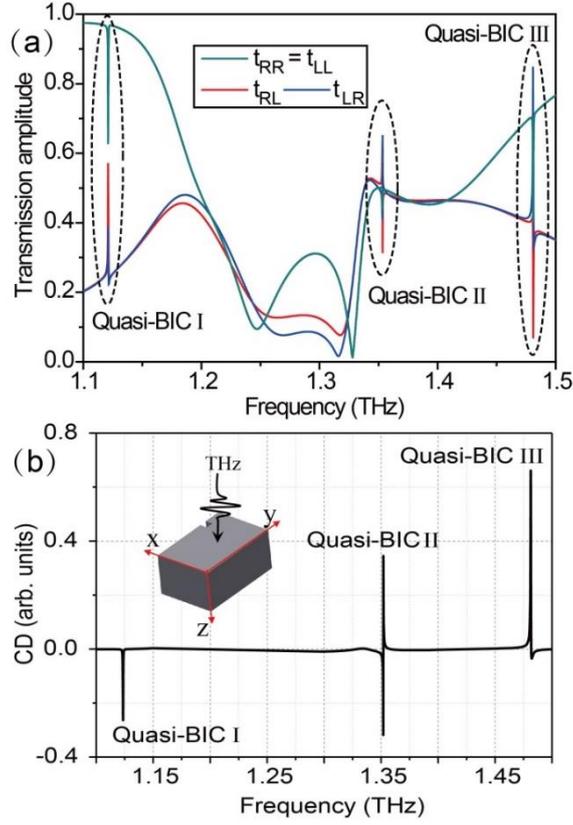

Fig. 2. (a) The transmission spectra of circular polarization waves for metasurface (g=12 μm) supporting chiral quasi-BIC, and (b) its circular dichroism (CD). It can be seen that three chiral quasi-BICs are shown.

To identify the BIC type, we first construct an in-plane mirror structure ($d=0$ μm, $g=12$ μm) that only breaks in-plane $C_2$ symmetry. Such structure provides convenience for the identification of symmetry-protected BIC. Fig. 3a-b show the optical band and the radiative Q factor for transverse electric (TE) modes in rectangular silicon without opening, and $TE_1$ mode at Γ point (1.09 THz) with infinite Q factor is found. The bound electromagnetic field with even symmetry is a magnetic dipole mode without out-plane radiation (Fig. 3b). Further, the transmission spectra depended on opening size are investigated in Fig. 3c-d. As the size of the central opening increases ($C_2$ symmetry breaking), the radiative Q factor (defined as $\omega_0/2\gamma_0$) decreases (Fig. 3d-e) with a relationship of $Q \propto \alpha^{-2}$ [14], where α describes the asymmetry degree and defined as the ratio of opening area to initial area (ΔS/S). These characteristics suggest that quasi-BIC I is a symmetry-protected type (BIC state is a TE mode). The second BIC is found in the results of the optical band and the radiative Q factor for transverse magnetic (TM) modes in rectangular silicon, as shown in Fig. 3f-g. The $TM_3$ mode at Γ point (1.34 THz) with infinite Q factor shows that bound electromagnetic field with even symmetry is a toroidal dipole mode without out-plane radiation (Fig. 3g). With the $C_2$ symmetry breaking, it can be seen from the transmission spectra that the radiative Q factor decreases with a relationship of $Q \propto \alpha^{-2}$, as shown in Fig. 3h-j. Obviously, quasi-BIC II is also assigned to the symmetry-protected type (BIC state is a TM mode).

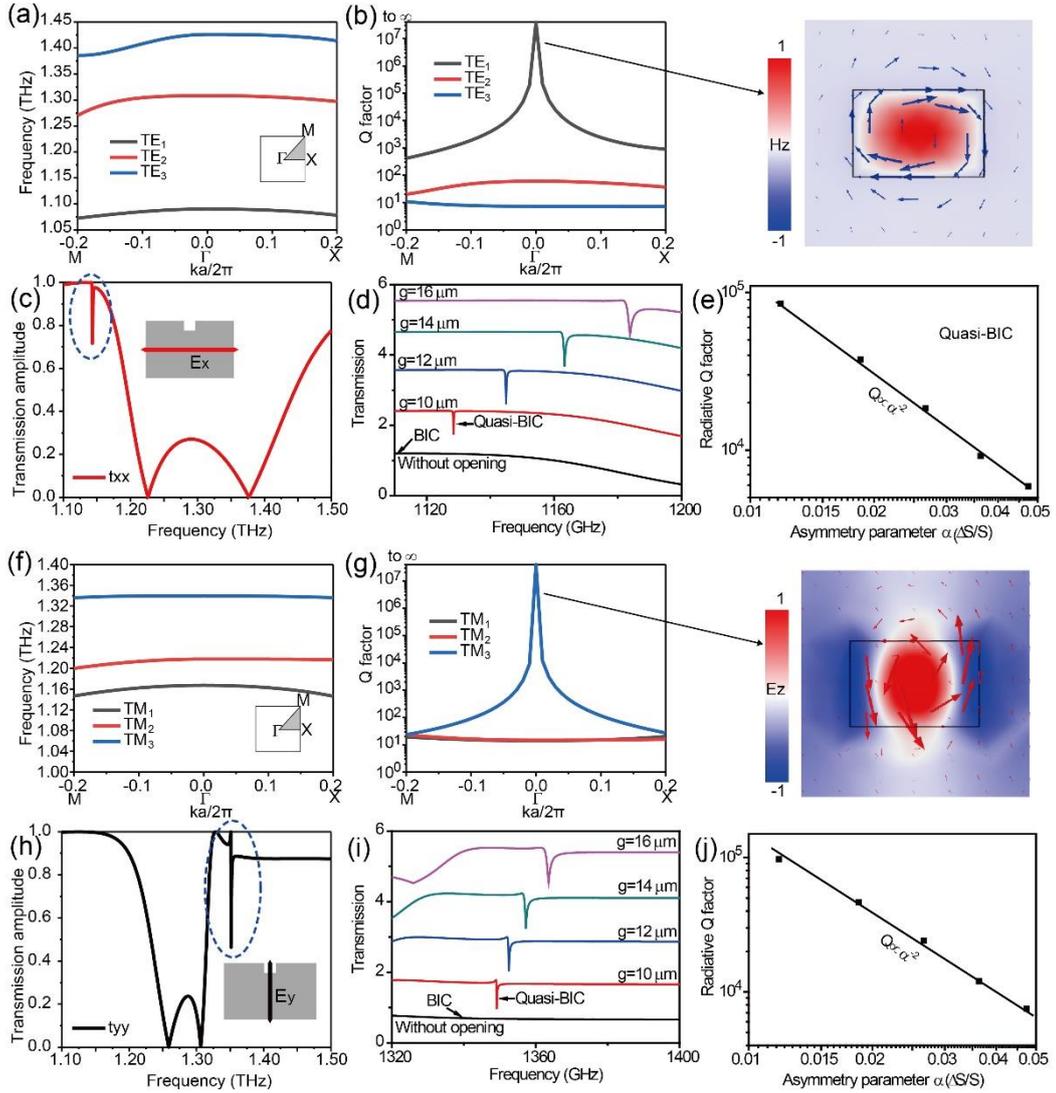

Fig. 3. (a-b) The optical band and the radiative Q factor for TE modes in rectangular silicon without opening, and inset in (b) means the normalized magnetic field distribution with z component at BIC state, where blue arrows mean electrical field line. (c) For the structure with d=0 μm, g=12 μm, the full spectra in the range of 1.1-1.5 THz under excitation of x-polarized wave. (d) The fine transmission spectra of the structure with different opening sizes, corresponding to (e) the radiative Q factor vs. α. (f-g) The optical band and the radiative Q factor for TM modes in rectangular silicon without opening, and inset in (g) means the normalized electrical field distribution with z component at BIC state, where red arrows mean magnetic field line. (h) For the structure with d=0 μm, g=12 μm, the full spectra in the range of 1.1-1.5 THz under excitation of y-polarized wave. (i) The fine transmission spectra of the structure with different opening sizes, corresponding to (j) the radiative Q factor vs. α, where coordinate value types are displayed as logarithmic plots.

In such structure with in-plane $C_2$ symmetry breaking but in-plane mirror symmetry remaining, a TM mode at Γ point (1.465 THz) with infinite Q factor is found, which corresponds to the third BIC, as shown in Fig. 4a-c. However, even if the $C_2$ symmetry has been destroyed, the third BIC located in the range of 1.45-1.5 THz is still not exposed in the far field. Obviously, the third BIC is not symmetry-protected type.

Further, by moving the opening position along x-axis, the structure loses the in-plane mirror symmetry and the quasi-BIC III is found, as shown in Fig. 4d-f. Again, the quasi-BIC III can enter into accidentally BIC state with the continuous adjustment of opening position to 35 μm (Fig. 4e-f). More evidences are shown in Fig. 4g-i to further identify the parameter-tuned BIC. Firstly, it can be seen from Fig. 4c that the annular electric field line is distributed in the y-z plane and the magnetic field is distributed in the x-y plane. A clearer schematic diagram is plotted in Fig. 4g, the electromagnetic field mode is essentially composed of two magnetic dipoles with opposite magnetic moments. Two magnetic moments are located in the x-y plane, and the electromagnetic field radiates out of the plane. The two magnetic dipoles have equal radiation intensity but a fixed phase difference of π (due to opposite magnetic moments labeled m1 and m2, respectively), so they produce completely cancelled radiation in each channel ($E_{m1}+E_{m2}=0$), which is the reason for the formation of zero radiation BIC (Fig. 4i). When the opening position changes, it can be seen from Fig. 4h the electric field line distribution for one of the magnetic dipoles has a slight deflection, which makes the two magnetic dipoles out of balance. As a result, the radiative vector sum of two magnetic dipoles is not zero ($E_{m1}+E_{m2}\neq0$), thus forming a quasi-BIC state (Fig. 4i). Such BIC, based on the coupling of all radiative waves in a resonance mode, belongs to the single resonance type [13]. Single resonance BIC is not guaranteed to exist by symmetry. The BIC without symmetry incompatibility also can give rise to a tunable trapping of light, which has been found by Chia Wei Hsu *et al.* in 2013 [26]. The coupled-wave theory explain usually the BIC where the single resonance itself is considered as destructive interference of several sets of radiative waves [27]. These radiative waves can be incorporated to cancel each other by tuning given parameter, thus achieving the overall radiation to be suppressed (close to zero radiation), which is unlike symmetry-protected BIC that zero radiation is caused by decoupling of bound state and continuum because of symmetry incompatibility.

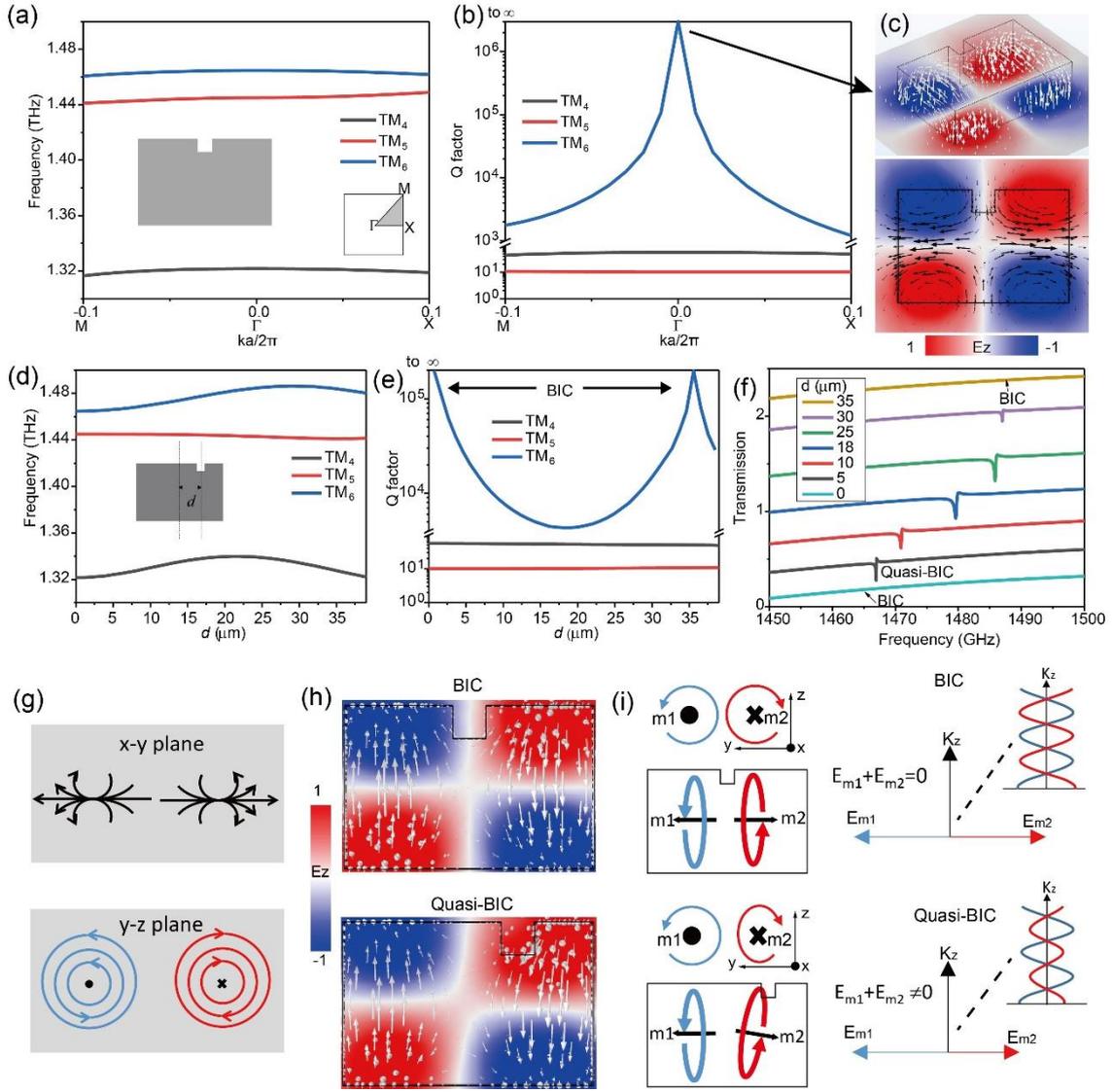

Fig. 4. (a-b) The optical band and the radiative Q factor for TM modes in the structure with in-plane $C_2$ symmetry breaking but in-plane mirror symmetry remaining. (c) The normalized electrical field distribution with z component at BIC state, where white arrows mean electrical field line and black arrows mean magnetic field line. (d-e) The eigenfrequency and the radiative Q factor of Γ point for TM modes in the structure with the change of opening position, corresponding to (f) the fine transmission spectra under the excitation of x-polarized wave. (g) A clear schematic picture for the magnetic field line (up) and electrical field line (down) at BIC state. (h) The electrical field changes with the opening position shifts, where white arrows mean electrical field line. (i) The schematic changes for two magnetic moments and out of plane radiation with the BIC transition to quasi-BIC, where blue and read arrows mean electrical field line, and black arrows means the magnetic moments.

Now, let's return to the discussion on chiral attribute of the metasurface. The quasi-BIC III is selected as a representative, and further investigate its fine reflection and transmission characteristics (obviously, the discussion on chiral attribute of the remaining two quasi-BICs does not lose generality, thus their spectral investigation is no longer given in detail here, see **S2** in **Supplementary Information**). Also, the opening size is reduced to 8 μm to obtain higher Q value with narrower linewidth. Considering the THz wave incident on port I, the detailed spectra are shown in Fig. 5. The transmission coefficients with Fano linetype are same for co-polarization waves ($t_{RR}=t_{LL}$) and the difference for the cross-polarization waves ($t_{RL} \neq t_{LR}$), as shown in Fig. 5a. The reflection coefficients with Fano linetype are same for the cross-polarization waves ($r_{RL}=r_{LR}$) and the difference for co-polarization wave ($r_{LL} \neq r_{RR}$), as shown in Fig. 5b. In our design, when RCP THz wave at quasi-BIC III is incident on metasurface from the top, most of them will be converted to LCP waves and transmitted (comparing $t_{RR}$ and $t_{LR}$ in Fig. 5a), with weak reflection (comparing $r_{RR}$ and $r_{LR}$ in Fig. 5b). However, when the LCP THz wave at the quasi-BIC frequency is incident on metasurface from the top, the co-polarization $t_{LL}$ and cross-polarization $t_{RL}$ components are barely transmitted (Fig. 5a), and most of them are reflected as co-polarization (comparing $r_{LL}$ and $r_{RL}$ in Fig. 5b). In short, the metasurface at the quasi-BIC III will mainly transmit RCP wave from the top and reflect LCP wave from the top. All simulative results in Fig. 5a-b are consistent with theoretical trends described in Eq. 1 by temporal CMT. Furthermore, Eq. 3 can fit perfectly the simulative results, and the real parameters $A_j$, $B_j$, $C_j$ and $D_j$ (j = 1, 2, 3, 4, 5, 6) are determined, as shown in Fig. 5c-h.

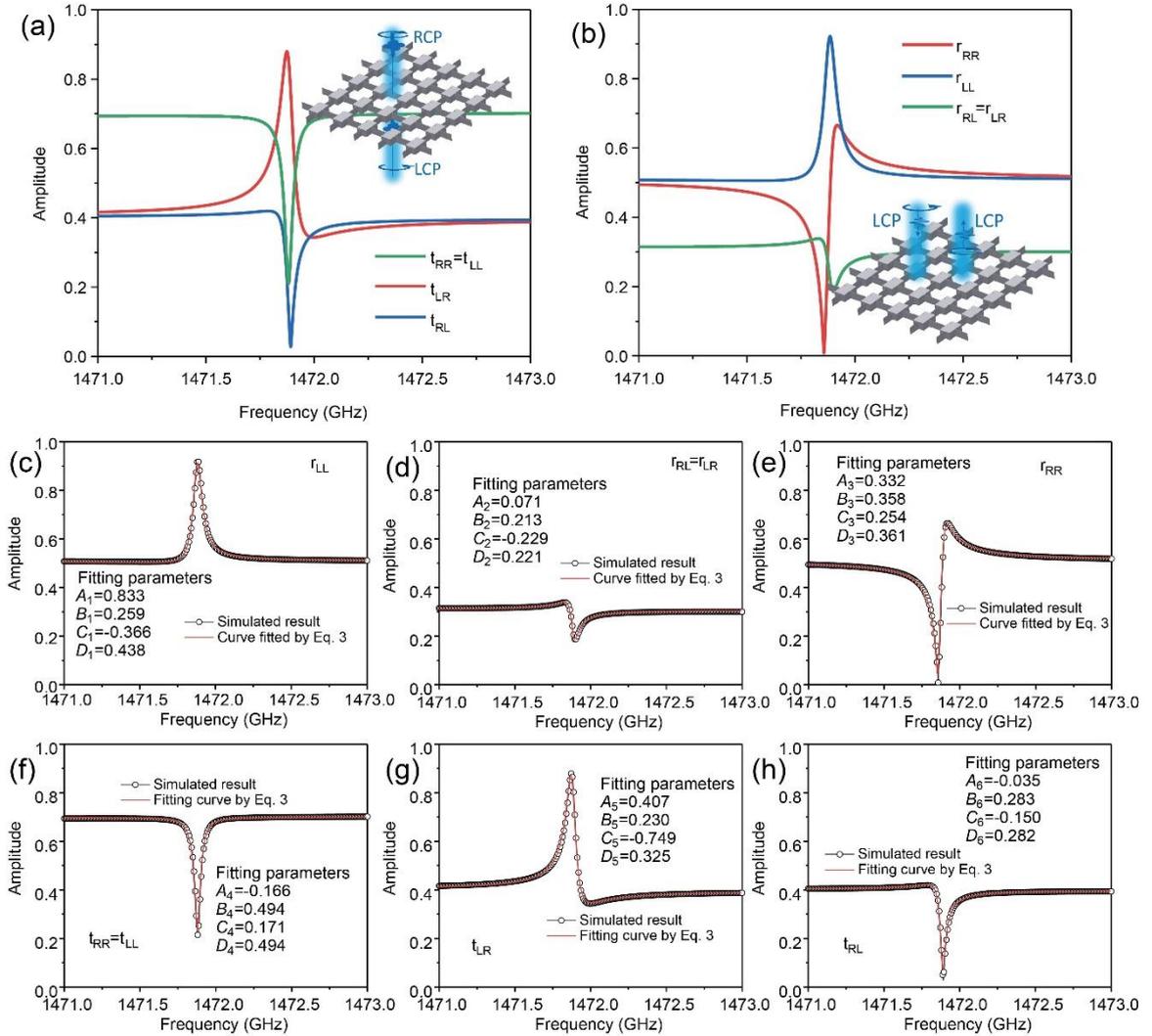

Fig. 5. For opening size of 8 μm, (a-b) respectively, transmission and refection amplitudes of circular polarization waves. (c-h) The comparison between simulated results and theoretical fitting results for (c) $r_{LL}$, (d) $r_{RL}= r_{LR}$, (e) $r_{RR}$, (f) $t_{RR}= t_{LL}$, (g) $t_{LR}$, and (h) $t_{RL}$.

Fig. 6a further shows the contrast of transmission circular dichroism of metasurfaces with different openings. Note that the quasi-BIC frequency shift as the opening size is due to changes in the structural asymmetry degree. When the opening size is 20 μm, the radiative Q factor has a magnitude of $10^2$, and the linewidth is less than 3 GHz. With the decrease of the opening size, the radiation leakage of quasi-BIC weakens. It can be seen that when the opening size is 12 μm, the radiation linewidth is less than 0.5 GHz with the radiative Q factor of $10^3$ level (Fig. 6b). Furthermore, when the opening size is 8 μm, the radiation linewidth is only 0.05-0.06 GHz with increasing radiative Q factor to the order of $10^4$ (Fig. 6c).

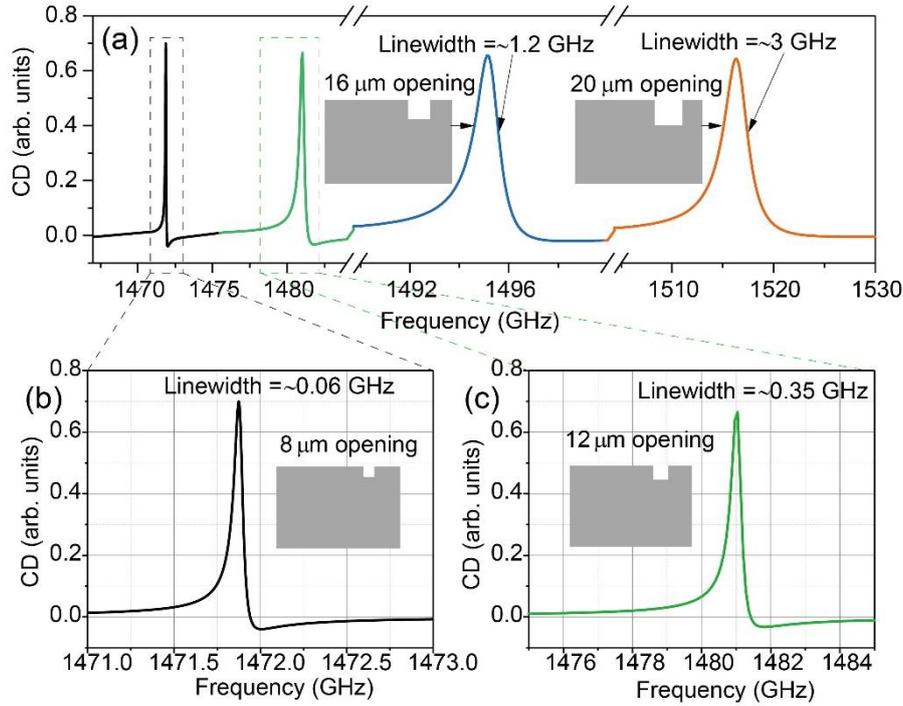

Fig. 6. (a) Transmission circular dichroism of metasurfaces with different openings, and (b-c) fine scanning for 8 μm opening and 12 μm opening.

## 4. CONCLUSION

In summary, we introduce chiral quasi-BIC mode into planar all-silicon metasurface, and realize ultra-narrowband THz chirality manipulation. Different from the past, although $C_2$ symmetry of the metasurface are broken, the mirror symmetry with respect to a plane perpendicular to the z-axis is still retained. This design is more conducive to the actual preparation of the device. Also, such system exposes several BICs including two symmetry-protected BICs and a parameter-tuned BIC assigned to single resonance type. The analytical expression of scattering matrix of chiral quasi-BIC response has been disclosed based on a temporal coupled-mode theory in this work, which is consistent with simulative trends. The results show that the THz chiral propagation is modulated by the frequency and the radiative loss of quasi-BIC mode. The frequency of circular polarization decoupling occurs at quasi-BIC frequency, and the radiative Q factor is as high as $10^4$ level with the radiation linewidth of only 0.05-0.06 GHz.

**Support Information**

Support Information are available.

**Acknowledgement**

This work was supported by the National Key Research and Development Program of China (Nos. 2021YFB2800703 and 2017YFA0700202), National Natural Science Foundation of China (No. 61675147), and Central Government Guided Local Science and Technology Development Projects, China (No.2021ZYD0039).

**Data Availability**

The data that support the findings of this study are available from the corresponding author upon reasonable request.

**Authorship contribution statement**

Jitao Li: Conceptualization, methodology and writing original manuscript. Zhen Yue: Methodology, Software. Jie Li and Chenglong Zheng: Data curation. Dingyu Yang, Silei Wang and Mengyao Li: Investigation. Yating Zhang and Jianquan Yao: Supervision, Reviewing.

**Conflict of interest statement**

The authors declare no conflict of interest.

# <Supplementary Information>

## S1. Scattering matrix of planar chiral quasi-BIC response

Fig. 1d in main text has shown that the radiative polarization state of at-Γ quasi-BIC possesses nonzero ellipticity. Under the framework of temporal coupled-mode theory (CMT), the dynamic relationship of resonance amplitude reads as:

$$\frac{d\mathbf{a}_c}{dt} = [i\omega_0 - \gamma_0]\mathbf{a}_c + \mathbf{D}^{\mathrm{T}}|I_{cir}\rangle, \tag{S1}$$

where $\mathbf{a}_c$ is a scalar amplitude which is defined by the energy stored in the eigenmode, and $\mathbf{D}$ describes the coupling between resonator and ports. The $|I_{cir}\rangle = \begin{bmatrix} I_R^{\mathrm{I}} & I_L^{\mathrm{I}} & I_R^{\mathrm{II}} & I_L^{\mathrm{II}} \end{bmatrix}^{\mathrm{T}}$ represent the incoming wave amplitudes of the RCP (or LCP) component from port I (or from port II). Next, we derive the scattering matrix of planar metasurface in the chiral quasi-BIC mode. First consider the outgoing wave $|O_{cir}^"\rangle$ of the metasurface under the coupling between modes. $|O_{cir}^"\rangle$ comes from two contributions. The first term comes from the light scattering caused by the ordinary achiral response $\mathbf{C}_{cir}|I_{cir}\rangle$, where coupling between the incoming wave and the outgoing wave through a direct pathway. The second term comes from the coupling between the resonant modes in the quasi-BIC case $|O_{cir}^{\mathbf{D}}\rangle = \mathbf{D}\mathbf{a}_c$. The final outgoing wave can be written as:

$$|O_{cir}^"\rangle = \mathbf{C}_{cir}|I_{cir}\rangle + \mathbf{D}\mathbf{a}_c = \mathbf{S}_{cir}^"|I_{cir}\rangle, \tag{S2}$$

where $|O_{cir}^"\rangle = \begin{bmatrix} O_R^{\mathrm{I}} & O_L^{\mathrm{I}} & O_R^{\mathrm{II}} & O_L^{\mathrm{II}} \end{bmatrix}^{\mathrm{T}}$ represent the outgoing wave amplitudes of the RCP (or LCP) component from port I (or from port II) after considering the coupling between modes, and $\mathbf{S}_{cir}^"$ is the scattering matrix in this case. It needs to be emphasized that in Eq. (S2) $\mathbf{C}_{cir}$ is direct scattering of metasurface with achiral, as shown in:

$$\mathbf{C}_{cir} = \begin{bmatrix} r_{RR}^{\mathrm{I}} & r_{RL}^{\mathrm{I}} & t_{RR}^{\mathrm{II}} & t_{RL}^{\mathrm{II}} \\ r_{LR}^{\mathrm{I}} & r_{LL}^{\mathrm{I}} & t_{LR}^{\mathrm{II}} & t_{LL}^{\mathrm{II}} \\ t_{RR}^{\mathrm{I}} & t_{RL}^{\mathrm{I}} & r_{RR}^{\mathrm{II}} & r_{RL}^{\mathrm{II}} \\ t_{LR}^{\mathrm{I}} & t_{LL}^{\mathrm{I}} & r_{LR}^{\mathrm{II}} & r_{LL}^{\mathrm{II}} \end{bmatrix} = \begin{bmatrix} r & r' & t & t' \\ r' & r & t' & t \\ t & t' & r & r' \\ t' & t & r' & r \end{bmatrix}, \tag{S3}$$

where $r, r', t, t'$ are complex numbers with the phase angle of $\exp(i\varphi_1)$, $\exp(i\varphi_2)$, $\exp(i\varphi_3)$ and $\exp(i\varphi_4)$ respectively. Also, there is $|r|^2 + |r'|^2 + |t|^2 + |t'|^2 = 1$. The $r_{jk}^{\mathrm{I/II}}$ and $t_{jk}^{\mathrm{I/II}}$ ( $j,k \in \{L,R\}$, L means LCP, and R means RCP), respectively mean reflection and transmission coefficients of j-component under k-component wave incident on port I/II. Considering the time reversal symmetry, the background scattering $\mathbf{C}_{cir}$ should be unitary, i.e. $\mathbf{C}_{cir}^{\dagger}\mathbf{C}_{cir} = \mathbf{I}$. In fact, the time reversal symmetry causes also a phase difference of π/2 between the transmission coefficient of a wave and the reflection coefficient of its handedness-flipped state [S1], namely $\varphi_3 = \varphi_2 + \pi/2$. Furthermore, for $|O_{cir}^{\mathbf{D}}\rangle = \mathbf{D}\mathbf{a}_c$, the eigenstate $\mathbf{a}_c$ is coupled with the circular polarization waves incident on the side of the metasurface with the coupling matrix $\mathbf{D}$, and the coupled waves $|O_{cir}^{\mathbf{D}}\rangle = [O_R^{D-\mathrm{I}} \; O_L^{D-\mathrm{I}} \; O_R^{D-\mathrm{II}} \; O_L^{D-\mathrm{II}}]^T$ output to the port, where $O_{L/R}^{D-\mathrm{I}/\mathrm{II}}$ means the coupling wave amplitudes of the outgoing LCP (or RCP) components at port I (or port II), and reads as:

$$\begin{bmatrix} O_R^{D-\mathrm{I}} \\ O_L^{D-\mathrm{I}} \\ O_R^{D-\mathrm{II}} \\ O_L^{D-\mathrm{II}} \end{bmatrix} = \begin{bmatrix} d_R^{\mathrm{I}} \\ d_L^{\mathrm{I}} \\ d_R^{\mathrm{II}} \\ d_L^{\mathrm{II}} \end{bmatrix} [a], \qquad (\text{S4a})$$

$$\mathbf{D} = \begin{bmatrix} d_R^{\mathrm{I}} \\ d_L^{\mathrm{I}} \\ d_R^{\mathrm{II}} \\ d_L^{\mathrm{II}} \end{bmatrix}. \qquad (\text{S4b})$$

where $d_j^{\mathrm{I}/\mathrm{II}}$ ($j \in \{L,R\}$) means the coupling coefficient where the resonant mode is coupled with the incoming wave at port I (or port II) and the output as $j$-component wave. Since the radiative rate is related to the coupling from resonance to the port, $\mathbf{D}$ matrix is not arbitrary. Also, $\mathbf{D}$ is limited by background scattering $\mathbf{C}_{cir}$. They satisfy the relationship [S2]:

$$\mathbf{D}^+\mathbf{D} = 2\gamma_0, \qquad (\text{S5a})$$

$$\mathbf{C}\mathbf{D}^* = -\mathbf{D}. \qquad (\text{S5b})$$

Next, we derive the specific form of the $\mathbf{D}$ matrix. It needs to be noted that the circular polarization in this system is not completely decoupled, which means that each eigenmode coupled to the port outputs both LCP and RCP. Due to the mirror symmetry of the system, the total amount of the outgoing LCP and RCP waves of an eigenmode coupled to the two ports is the same, namely:

$$d_L^{\mathrm{I}} + d_R^{\mathrm{I}} = d_L^{\mathrm{II}} + d_R^{\mathrm{II}}. \qquad (\text{S6})$$

In addition, the outgoing waves with cross-polarization of an eigenmode coupled to the two ports should be the same coupling coefficient, we have:

$$\begin{cases} d_L^{\mathrm{I}} = d_R^{\mathrm{II}} = m \\ d_R^{\mathrm{I}} = d_L^{\mathrm{II}} = n \end{cases}, \qquad (\text{S7})$$

Thus $\mathbf{D}$ matrix reads as:

$$\mathbf{D} = \begin{bmatrix} m \\ n \\ n \\ m \end{bmatrix}. \qquad (\text{S8})$$

Due to the coupling with phase angle, the elements in Eq. (S8) are complex. According to the Eq. (S5a), we can find the expression of each element of the $\mathbf{D}$ matrix as follows:

$$|m|^2 + |n|^2 = \gamma_0. \qquad (\text{S9})$$

Solving Eq. (S9) can obtain:

$$\begin{cases} m = \exp(i\theta_m)\sqrt{\beta\gamma_0} \\ n = \exp(i\theta_n)\sqrt{\alpha\gamma_0} \end{cases}, \qquad (\text{S10})$$

where $0 < \beta \neq 0.5 < 1$ and $\beta + \alpha = 1$. The $\exp(i\theta)$ is the coupled phase angle. The $\beta$ and $\alpha$ essentially relate to the ratio of LCP and RCP waves of eigenmode coupled to ports. For instance, with mode coupled to port

I, outgoing RCP wave amplitude ratio of is $\sqrt{\beta}$ and outgoing LCP wave amplitude is $\sqrt{\alpha}$, so that the total intensity of outgoing waves is $(\sqrt{\beta})^2 + (\sqrt{\alpha})^2 = 1$.

Furthermore, by combining Eqs. (S1) and (S2), the expression $\mathbf{S}_{cir}^{"}$ can be obtained:

$$\mathbf{S}_{cir}^{"} = \mathbf{C}_{cir} + \frac{\mathbf{D}\mathbf{D}^T}{i(\omega-\omega_0)+\gamma_0}. \tag{S11}$$

Combining Eqs. (S3) and (S8) into Eq. (S11) can get:

$$\mathbf{S}_{cir}^{"} = \begin{bmatrix} r_{RR}^I & r_{RL}^I & t_{RR}^{II} & t_{RL}^{II} \\ r_{LR}^I & r_{LL}^I & t_{LR}^{II} & t_{LL}^{II} \\ t_{RR}^I & t_{RL}^I & r_{RR}^{II} & r_{RL}^{II} \\ t_{LR}^I & t_{LL}^I & r_{LR}^{II} & r_{LL}^{II} \end{bmatrix} = \begin{bmatrix} r + \frac{m^2}{i(\omega-\omega_0)+\gamma_0} & r' + \frac{mn}{i(\omega-\omega_0)+\gamma_0} & t + \frac{mn}{i(\omega-\omega_0)+\gamma_0} & t' + \frac{m^2}{i(\omega-\omega_0)+\gamma_0} \\ r' + \frac{mn}{i(\omega-\omega_0)+\gamma_0} & r + \frac{n^2}{i(\omega-\omega_0)+\gamma_0} & t' + \frac{n^2}{i(\omega-\omega_0)+\gamma_0} & t + \frac{mn}{i(\omega-\omega_0)+\gamma_0} \\ t + \frac{mn}{i(\omega-\omega_0)+\gamma_0} & t' + \frac{n^2}{i(\omega-\omega_0)+\gamma_0} & r + \frac{n^2}{i(\omega-\omega_0)+\gamma_0} & r' + \frac{mn}{i(\omega-\omega_0)+\gamma_0} \\ t' + \frac{m^2}{i(\omega-\omega_0)+\gamma_0} & t + \frac{mn}{i(\omega-\omega_0)+\gamma_0} & r' + \frac{mn}{i(\omega-\omega_0)+\gamma_0} & r + \frac{m^2}{i(\omega-\omega_0)+\gamma_0} \end{bmatrix}, \tag{S12}$$

Through a series of straightforward but quite cumbersome math operations of Eq. S12, the reflection and transmission coefficients can be simplified as a form of Fano linetype function with a center frequency $\omega_0$ and a radiative rate $\gamma_0$, and written as a form of: $(\Lambda+i\Gamma)/[i(\omega-\omega_0)+\gamma_0]$, where $\Lambda$, $\Gamma$ are real numbers, as follows:

$$\mathbf{S}_{cir}^{"} = \begin{bmatrix} r_{RR}^I & r_{RL}^I & t_{RR}^{II} & t_{RL}^{II} \\ r_{LR}^I & r_{LL}^I & t_{LR}^{II} & t_{LL}^{II} \\ t_{RR}^I & t_{RL}^I & r_{RR}^{II} & r_{RL}^{II} \\ t_{LR}^I & t_{LL}^I & r_{LR}^{II} & r_{LL}^{II} \end{bmatrix} = \begin{bmatrix} \frac{\Lambda_1+i\Gamma_1}{i(\omega-\omega_0)+\gamma_0} & \frac{\Lambda_2+i\Gamma_2}{i(\omega-\omega_0)+\gamma_0} & \frac{\Lambda_4+i\Gamma_4}{i(\omega-\omega_0)+\gamma_0} & \frac{\Lambda_6+i\Gamma_6}{i(\omega-\omega_0)+\gamma_0} \\ \frac{\Lambda_2+i\Gamma_2}{i(\omega-\omega_0)+\gamma_0} & \frac{\Lambda_3+i\Gamma_3}{i(\omega-\omega_0)+\gamma_0} & \frac{\Lambda_5+i\Gamma_5}{i(\omega-\omega_0)+\gamma_0} & \frac{\Lambda_4+i\Gamma_4}{i(\omega-\omega_0)+\gamma_0} \\ \frac{\Lambda_4+i\Gamma_4}{i(\omega-\omega_0)+\gamma_0} & \frac{\Lambda_5+i\Gamma_5}{i(\omega-\omega_0)+\gamma_0} & \frac{\Lambda_3+i\Gamma_3}{i(\omega-\omega_0)+\gamma_0} & \frac{\Lambda_2+i\Gamma_2}{i(\omega-\omega_0)+\gamma_0} \\ \frac{\Lambda_6+i\Gamma_6}{i(\omega-\omega_0)+\gamma_0} & \frac{\Lambda_4+i\Gamma_4}{i(\omega-\omega_0)+\gamma_0} & \frac{\Lambda_2+i\Gamma_2}{i(\omega-\omega_0)+\gamma_0} & \frac{\Lambda_1+i\Gamma_1}{i(\omega-\omega_0)+\gamma_0} \end{bmatrix}, \tag{S13}$$

where

$$\Lambda_j + i\Gamma_j = [\gamma_0 A_j + (\omega-\omega_0)B_j] + i[\gamma_0 C_j + (\omega-\omega_0)D_j], j \in \{1,2,3,4,5,6\}, \tag{S14}$$

in detail, there are:

$$\begin{cases} A_1 = |r|\cos\varphi_1 + \beta\cos 2\theta_m, B_1 = -|r|\sin\varphi_1, C_1 = |r|\sin\varphi_1\beta\sin 2\theta_m, D_1 = |r|\cos\varphi_1 \\ A_2 = |r'|\cos\varphi_2 + \sqrt{\beta\alpha}\cos(\theta_m+\theta_n), B_1 = -|r'|\sin\varphi_2, C_2 = |r'|\sin\varphi_2 + \sqrt{\beta\alpha}\sin(\theta_m+\theta_n), D_2 = |r'|\cos\varphi_2 \\ A_3 = |r|\cos\varphi_1 + \alpha\cos 2\theta_n, B_3 = -|r|\sin\varphi_1, C_3 = |r|\sin\varphi_1 + \alpha\sin 2\theta_n, D_3 = |r|\cos\varphi_1 \\ A_4 = |t|\cos\varphi_3 + \sqrt{\beta\alpha}\cos(\theta_m+\theta_n), B_4 = -|t|\sin\varphi_3, C_4 = |t|\sin\varphi_3 + \sqrt{\beta\alpha}\sin(\theta_m+\theta_n), D_4 = |t|\cos\varphi_3 \\ A_5 = |t'|\cos\varphi_4 + \alpha\cos 2\theta_n, B_5 = -|t'|\sin\varphi_4, C_4 = |t'|\sin\varphi_4 + \alpha\sin 2\theta_n, D_4 = |t'|\cos\varphi_4 \\ A_6 = |t'|\cos\varphi_4 + \beta\cos 2\theta_m, B_6 = -|t'|\sin\varphi_4, C_4 = |t'|\sin\varphi_4 + \beta\sin 2\theta_m, D_4 = |t'|\cos\varphi_4 \end{cases}. \tag{S15}$$

Although the expression for $A_j$, $B_j$, $C_j$ and $D_j$ looks quite complex in Eq. S15, they will be determined when the structure is determined. In Eq. S13, the frequency and the radiative loss of all components are corresponded to the frequency and the radiative loss of quasi-BIC. By adjusting the Q factor of quasi-BIC, the $\gamma_0$ can be reduced to an extremely low level.

### S2. Transmission spectra and theoretical fitting results for chiral quasi-BICs I and II

Fig. S1(a) and S1(c) show the fine transmission spectra for chiral quasi-BICs I and II, respectively. Without losing generality, Eq. 3 in main text can also fit perfectly the simulative results of quasi-BIC I and II, as shown in Fig. S1(b) and S1(d). Here, the $t_{RR}$ is selected as a representative to prove the generality of theoretical fitting.

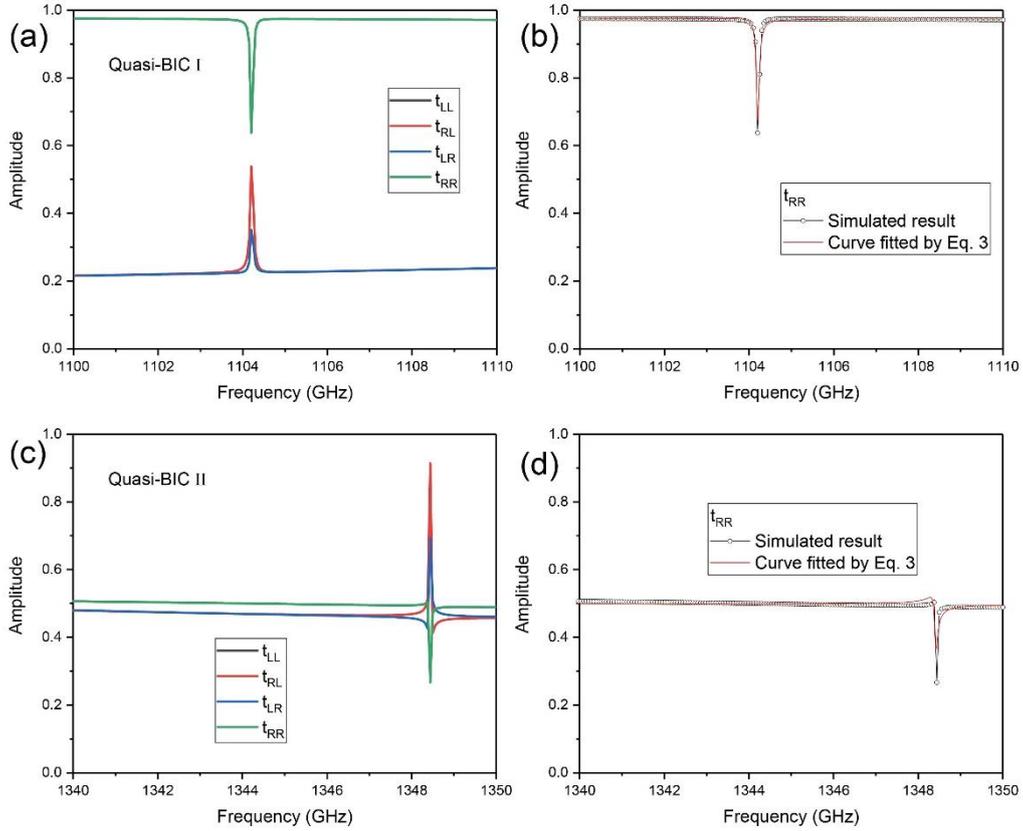

Fig. S1. (a, c) The fine transmission spectra for chiral quasi-BICs I and II, respectively. Correspondingly, (b, d) the comparison between simulated result and theoretical fitting result.